\documentclass[10pt]{article}
\usepackage{OSAmeet}
\usepackage{amssymb}
\newcommand{\abs}[1]{\left| #1 \right|}
\newcommand{\ket}[1]{\left| #1 \right\rangle}
\begin{document}

\title{Topological decomposition of composite quantum state spaces}

\author{Scott N. Walck}
\address{Department of Physics, Lebanon Valley College,
         Annville, PA 17003}
\email{717-867-6153, fax 717-867-6075, walck@lvc.edu}

\HeaderAuthorTitleMtg{Walck}{Topological decomposition...}{ICQI 2001}

\begin{abstract}
We present a two-part program for state space decomposition.
States are classified into entanglement classes based on
local unitary transformations, and then characterized
as elements of topological spaces using the language of fibre bundles.
\end{abstract}

\noindent                      
Much effort has been spent in attempting to assign a single
real number to a quantum state as a measure of entanglement.
\cite{vedral97,hill97,wootters98}
Here we take precisely the opposite approach.
We desire a description of every quantum state a system
can possess in a way that makes the entanglement properties
of the state explicit.  We want to classify and characterize quantum
states based on entanglement.
Since there is a notion of ``closeness'' of quantum
states that comes from transition probabilities between
states, we focus on the topological structure of the state space.

Our general scheme for classification of quantum states is based
on the following property of composite quantum systems.  Not all
states of a composite quantum system can be obtained by performing
local unitary operations on the system.
Rather, the group of local state transformations $G$ acts on the
composite system state space $S$ to partition the state space into orbits.
(The group $G$ is the group of local unitary transformations
with redundancy removed.  For example, $G = SO(3)$ for a single
qubit rather than $SU(2)$ since the latter has a two-fold redundancy.)
These orbits are equivalence classes of states with
two states regarded as equivalent if they are related by a
local state transformation.
Each equivalence class can be regarded as an entanglement class.
Under the quotient topology, the set of orbits
becomes the space of entanglement classes $\tilde{S}$.
There is a natural map $p: S \to \tilde{S}$ which associates
with each state its entanglement class.

That was the classification program.  The second part
is a full characterization of each quantum state starting with
its entanglement class.  What additional information is required?
We must come up with a way to differentiate states related by
a local state transformation.
For this it is convenient to pick a standard state in each
entanglement class, and give the local state transformation
which takes the standard state into the desired state.
So, for each class in $\tilde{S}$ we pick a state in $S$ to be
the standard representative of that entanglement class.
This gives us a map $f: \tilde{S} \to S$.  Naturally, this map must be
such that $p \circ f = i_{\tilde{S}}$,
where $i_{\tilde{S}}$ is the identity map on
$\tilde{S}$.
In general, there will be many distinct local state transformations
which take the same standard state into the same desired state,
so we also need a way to get rid of this redundancy.
Getting rid of this redundancy is a different job for different
entanglement classes.
For each class $e$ in $\tilde{S}$ we proceed as follows.
The characterization problem amounts to finding a description
for the space of states $p^{-1}(e)$.  (The topology for
this space is the subspace topology inherited from $S$.)
These are all of the states
in the same entanglement class as $e$.
We identify the stabilizer, or little group, of $f(e)$, $L_{f(e)}$.
This is the subgroup of $G$ which maps $f(e) \in S$
to itself in the action of $G$ on $S$.
The space of cosets $G/L_{f(e)}$ is then homeomorphic to
$p^{-1}(e)$.

In principle, this program could be carried out for any
composite quantum state space, including spaces of pure
states or mixed states.

As an example, we look at both the classification
and characterization parts of the program for the simplest situation:
pure states of a pair of qubits.
Here, the state space $S = \mathbb{CP}^3$, the complex projective
space of three complex dimensions (six real dimensions) formed as
the projective space of the Hilbert space $\mathbb{C}^4$ by
removing the physically unimportant overall amplitude and phase.
The group of local state transformations $G = SO(3) \times SO(3)$.
We can view these transformations as pairs of rotations, one for each
qubit.
The space of entanglement classes is the closed unit interval,
$\tilde{S} = [0,1]$.  This comes about because of the Schmidt decomposition
theorem\cite{ekert95,aravind96},
and says that the single real number measure of
entanglement is complete for pure states of a pair of qubits.
We take this single real number measure of entanglement
to be the concurrence\cite{hill97,wootters98}, $C$.
For an arbitrary pure state,
\begin{equation}
\label{eqpurestate}
c_{++} \ket{++} + c_{+-} \ket{+-} + c_{-+} \ket{-+} + c_{--} \ket{--},
\end{equation}
the concurrence is given by
\begin{equation}
\label{eqconc}
C = 2 \abs{c_{++} c_{--} - c_{+-} c_{-+}} = \sin \eta .
\end{equation}
Equation (\ref{eqconc}) effectively gives the map $p: S \to \tilde{S}$, and
has also been used to define the angle
$\eta$, which is an alternate way of specifying the entanglement.
The end points $C=0$ and $C=1$ correspond to unentangled and fully
entangled states, respectively, and the points in between represent
partially entangled states.

Now, let's look at the characterization part.  For each class $C \in [0,1]$,
(or, equivalently, for each $\eta \in [0,\pi/2]$)
we identify a Schmidt standard state
$\ket{\psi_0(\eta)} = \cos(\eta/2) \ket{++} + \sin(\eta/2) \ket{--}$.
Next, we identify the little group that leaves the Schmidt standard
state invariant.  At this point we need to treat three separate cases:
the unentangled states ($C=0$), the partially entangled states
($0 < C < 1$), and the fully entangled states ($C = 1$).

For the unentangled states, we find that rotations about the $z$-axis
for either qubit leave the standard state invariant, and so are members
of the little group $SO(2) \times SO(2)$.
The unentangled state space $p^{-1}(0) = S^2 \times S^2$.
This is two Bloch spheres, one for each qubit, which
is what we expect in the absence of entanglement.

We treat the fully entangled states next.
Here, we find that for any rotation of qubit 1, there is a corresponding
rotation of qubit 2 that leaves the standard state invariant.
Consequently, the little group in this case is $SO(3)$.
The fully entangled state space is
$p^{-1}(1) = \mathbb{RP}^3 \approx SO(3)$.
This means that the space of fully entangled states is
homeomorphic to the three-dimensional rotation group, which
has the same topology as three-dimensional real projective space
$\mathbb{RP}^3$, the space of lines through the origin in $\mathbb{R}^4$.
So, there is a one-to-one correspondence between
three-dimensional rotations and fully entangled states.
For example, we can take the fully entangled
state associated with a counterclockwise rotation by
an angle $\phi$ about a unit vector $\hat{\mathbf{n}}$ to be
\[
e^{-i \hat{\mathbf{n}} \cdot \vec{\sigma} \phi / 2} \otimes I
\left| \mbox{singlet} \right\rangle
\]
where $\vec{\sigma}$ is the vector of Pauli matrices and $I$ is
the identity operator on the second qubit.
This description of fully entangled states has some nice properties.
First, the four Bell basis states are the singlet and the fully entangled
states associated with $\pi$ rotations about the
$x$-, $y$-, and $z$-directions.
For example, the Bell basis state $1/\sqrt{2} \ket{++} - 1/\sqrt{2} \ket{--}$
is associated with a $\pi$ rotation about the $x$-axis.
Second, a state associated with a rotation about $\hat{\mathbf{n}}$
is invariant to real rotations about that axis carried out identically
on both
subsystems.  For example, the Bell basis state above is left
invariant by any rotations about the $x$-axis applied identically
to both qubits.
The singlet state has a special role and is associated with the
identity in $SO(3)$.  The singlet is the only fully entangled state
(and the only pure state of two qubits) to be invariant under
\emph{all} rotations performed identically on both qubits.

Let us consider now the partially entangled states.
Here we find that rotations about the $z$-axis for qubit 1
accompanied by rotations about the $z$-axis for qubit 2
by the same angle in the \emph{opposite} direction
leave the standard state invariant, so that the little group
is $SO(2)$.
The space of partially entangled states $p^{-1}(C)$
with a fixed concurrence $C \in (0,1)$ is
a fibre bundle\cite{nakahara90}
made up of a base space $S^2 \times S^2$ and a fibre $S^1$.
The base space coordinates are
$(\theta_1,\phi_1)$ and $(\theta_2,\phi_2)$.
There are four different fibre coordinates (only one of which
is required in a given situation), and a corresponding collection
of four overlapping coordinate domains where each fibre coordinate
is valid.
The four fibre coordinates are $\gamma_{ij}$ and the
coordinate domains
are $U_{ij} = U_i \times U_j$, where $i$ and $j$
each run over N (for north) and S (for south).
Using Cartesian coordinates for the points on a sphere,
$U_N = S^2 - (0,0,-1)$ is the sphere with the south pole deleted and
$U_S = S^2 - (0,0,1)$ is the sphere with the north pole deleted.
Transition functions relate the four different $\gamma_{ij}$'s
on the regions where the coordinate domains overlap.
\begin{eqnarray}
t_{NN,SN} &=& t_{NS,SS} = e^{i 2 \phi_1} \\
t_{NN,NS} &=& t_{SN,SS} = e^{i 2 \phi_2} \\
t_{NN,SS} &=& e^{i 2 \phi_1} e^{i 2 \phi_2} \\
t_{NS,SN} &=& e^{i 2 \phi_1} e^{-i 2 \phi_2}
\end{eqnarray}
For example,
$e^{i \gamma_{NN}} = t_{NN,SN} e^{i \gamma_{SN}}
                   = e^{i 2 \phi_1} e^{i \gamma_{SN}}$.
These transition functions take values in a structure group $U(1) \approx S^1$
which maps the fibre onto itself.
Because the structure group and the fibre are the same space
$U(1) \approx S^1$, this fibre bundle is a principal bundle.
Using the shorthand operators,
\begin{eqnarray}
\mathsf{T}_N(\theta,\phi) &=&
e^{-i \sigma_z \phi / 2} e^{-i \sigma_y \theta / 2}
e^{i \sigma_z \phi / 2} \\
\mathsf{T}_S(\theta,\phi) &=&
e^{-i \sigma_z \phi / 2} e^{-i \sigma_y (\theta-\pi) / 2}
e^{i \sigma_z \phi / 2}
\end{eqnarray}
the kets corresponding to our parameters are given for base space
coordinates in
$U_{NN}$, $U_{NS}$, $U_{SN}$, and $U_{SS}$, respectively, as
\begin{equation}
\mathsf{T}_N(\theta_1,\phi_1) e^{-i \sigma_z \gamma_{NN} / 2}
\otimes
\mathsf{T}_N(\theta_2,\phi_2) \ket{\psi_0(\eta)}
\end{equation}
\begin{equation}
\mathsf{T}_N(\theta_1,\phi_1) e^{-i \sigma_z \gamma_{NS} / 2}
\otimes
\mathsf{T}_S(\theta_2,\phi_2) e^{-i \sigma_y \pi / 2}
\ket{\psi_0(\eta)}
\end{equation}
\begin{equation}
\mathsf{T}_S(\theta_1,\phi_1) e^{i \sigma_z \gamma_{SN} / 2}
e^{-i \sigma_y \pi / 2}
\otimes
\mathsf{T}_N(\theta_2,\phi_2) \ket{\psi_0(\eta)}
\end{equation}
\begin{equation}
\mathsf{T}_S(\theta_1,\phi_1) e^{i \sigma_z \gamma_{SS} / 2}
e^{-i \sigma_y \pi / 2}
\otimes
\mathsf{T}_S(\theta_2,\phi_2) e^{-i \sigma_y \pi / 2}
\ket{\psi_0(\eta)}
\end{equation}

The complete decomposition for pure states of two qubits is shown
in Fig. 1.
In general, one needs six parameters to characterize a
pure state of two qubits.  In our scheme, these are the concurrence
$C$ and five parameters used to pick out a point in the
$S^1$ bundle over $S^2 \times S^2$.  In the special cases of
unentangled states and fully entangled states, only four and three
parameters are needed, respectively.

The decomposition presented here lends itself well to a visualization
of quantum state space dynamics.  To visualize dynamics, one wants
a way to represent a quantum state which is both unique
(the Hilbert space description of pure states suffers from
overall amplitude and phase ambiguity) and continuous (so that
the system does not appear to make discontinuous jumps during
continuous evolution).  This decomposition satisfies both of these
criteria (with the minor annoyance of the multiple coordinate
systems from the fibre bundle)
while explicitly displaying the entanglement information.

It would be very interesting to know what the space of entanglement
classes is for larger composite systems.  In particular, the situation
for pure states of three qubits and mixed states of two qubits could
shed light on the nature of entanglement and give clues about
entanglement in more complex systems.

\begin{figure}
\begin{center}
\input{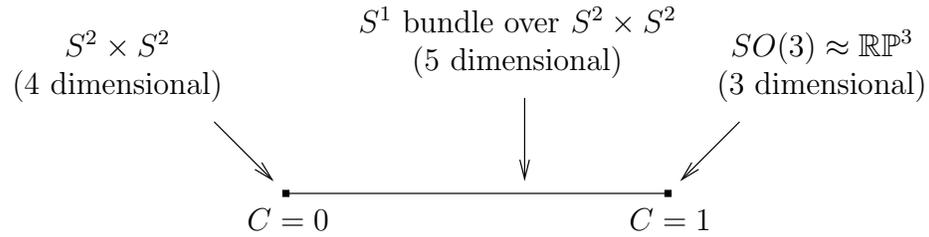}
\caption{Decomposition of $\mathbb{CP}^3$, the
space of pure states of two qubits.  The closed unit interval $[0,1]$ is
the space of entanglement classes.  The space of unentangled states
is homeomorphic to $S^2 \times S^2$.  The space of partially entangled
states with a fixed concurrence $C \in (0,1)$ is homeomorphic to an
$S^1$ bundle over $S^2 \times S^2$.  The space of fully entangled states
is homeomorphic to $SO(3) \approx \mathbb{RP}^3$.
}
\end{center}
\end{figure}

I thank Nathan Hansell, David Lyons, Chris Brazfield,
Pat Brewer, and Mike Fry for useful discussions.

\end{document}